\documentstyle[aps,prl,floats,graphicx,epsfig]{revtex}
\begin{document}
\draft
\twocolumn[\hsize\linewidth\columnwidth\hsize\csname @twocolumnfalse\endcsname

\title
{\bf Flow instability in 3He-A as analog of generation of hypermagnetic
field in early Universe.}
\author{
M. Krusius$^{1}$,   T. Vachaspati$^{2}$
and G.E. Volovik$^{1,3}$ }

\address{ $^1$ Low Temperature Laboratory, Helsinki University of
Technology, 02015 Espoo, Finland\\
$^2$ Physics Department, Case Western Reserve University,
Cleveland, OH 44106, USA\\
$^3$ Landau Institute for Theoretical Physics, 117334
Moscow, Russia. }

\date{\today} \maketitle

\
]
{\bf
It is now well-recognized that the Universe may behave like a
condensed matter system in which several phase transitions
have taken place. Superconductors and the superfluid phases of $^3$He are
condensed matter systems with useful similarities to the Universe: they both
contain Bose fields (order parameter) and Fermions (quasiparticles)
which interact in a way similar to the interaction of Higgs and gauge
particles with fermions in particle physics \cite{exotic,VolovikVachaspati}.
This analogy allows us to simulate many properties of the cosmologically
relevant physical (particle physics) vacuum in condensed matter, while
direct experiments in particle physics are still far from realization.
Recently, the anomalous generation of momentum (called ``momentogenesis'')
was experimentally confirmed in $^3$He: in the non-trivial background of
a moving $^3$He vortex, quantum effects gave rise to the production of
quasiparticles with momentum which were detected by measuring the force
on the vortex \cite{BevanNature}.
This phenomenon is based on the same physics as the anomalous generation of
matter in particle physics and bears directly on the cosmological problem of
why the Universe contains much more matter than antimatter (``baryogenesis'').
Here we report the experimental observation of the effect opposite to
momentogenesis: the conversion of quasiparticle momentum into a non-trivial
order parameter configuration or ``texture''. The corresponding process in a
cosmological setting would be the creation of a primordial magnetic field due
to changes in the matter content.
}

Processes in which magnetic fields are generated are very relevant to
cosmology as magnetic fields are ubiquitous in the universe. The Milky
Way, other galaxies, and clusters of galaxies are observed to have a
magnetic field whose generation is still not fully understood. One possible
mechanism is that a seed field was amplified by the complex motions
associated with galaxies and clusters of galaxies. The seed field itself
is usually assumed to be of cosmological origin.

In earlier work, it has been noted that the two genesis problems in
cosmology -- baryo- and magneto-genesis -- may be related to each other
\cite{roberge,tvgf,tvsintra}. More recently, a stronger possible connection
has been detailed in \cite{JoyceShaposhnikov,GiovanniniShaposhnikov}
(related work may be found in \cite{Vilenkin}). In this work, magnetogenesis
proceeds in the following two steps:

\noindent (i) At an early stage of the universe, possibly at the
Grand Unification epoch ($10^{-35}$ s after the big bang), an excess of
chiral right-handed electrons,
$e_R$, is produced due to parity violation. This is effectively
described by introducing a chemical potential, $\mu_R > 0$, for the
right-handed electrons leading to a certain number of $e_R$ in thermal
equilibrium at temperature $T$. The equilibrium relativistic energy and
particle number density are:
\begin{equation}
\epsilon_R = {1\over 6} T^2 \mu_R^2 ~~,~~n_R ={\partial \epsilon_R\over
\partial \mu_R}= {1\over 3} T^2
\mu_R
\label{ernr}
\end{equation}

\noindent (ii)
A striking property of a system with chiral fermions is that fermionic
charge such as $(n_R-n_L)$ -- the number of right-handed minus the number of
left-handed particles -- is conserved at the classical level but not if
quantum properties of the physical vacuum are taken into account. This charge
can be transferred to the ``inhomogeneity'' of the vacuum via the axial
anomaly as was first theoretically  predicted by  Adler \cite{Adler1969}
and Bell and Jackiw\cite{BellJackiw1969}, while its condensed matter analogue
was recently verified experimentally in $^3$He \cite{BevanNature}.
The inhomogeneity which absorbs the fermionic charge arises as a
magnetic field configuration, and the charge absorbed by the magnetic field,
${\bf\nabla}\times {\bf A}$, can be expressed as
\begin{equation}
(n_R-n_L)_{\bf A}={1\over 2\pi^2} {\bf A}\cdot ({\bf\nabla}\times {\bf A})~~.
\label{anomaly}
\end{equation}
The right-hand side is the so called Chern-Simons (or topological) charge
of the magnetic field.

The transformation of particles into a magnetic field configuration opens the
possibility for the cosmological origin of a magnetic field from a system
of fermions and this is a key step in the scenario described by
Joyce and Shaposhnikov \cite{JoyceShaposhnikov}. The $e_R$ excess generated
in the early universe survives until the electroweak phase transition (at about
$10^{-10}$ s after the big bang) when anomalous lepton (and baryon) number
violating processes become efficient and can erase the excess.
In doing so, there is an instability towards the production of a hypermagnetic
field. Since, a part of the hypermagnetic field is the electromagnetic field,
the present universe contains a primordial (electromagnetic) magnetic field.

Now we discuss the corresponding process in $^3$He-A, in which
quasiparticle momentum (the relative flow of normal and superfluid
components) is transformed via the chiral anomaly into the order parameter
texture. This is, in essence, the counterflow instability in $^3$He-A, which
has been intensively discussed theoretically
(see \cite{ThesisVollhardt} and Section 7.10 of the book
\cite{VollhardtWolfle})
and recently investigated experimentally in the Helsinki rotating cryostat
\cite{Experiment}. The $^3$He-A analogy of the cosmological scenario
described in
\cite{JoyceShaposhnikov} closely follows the two steps outlined
above.

\begin{figure}[!!!t]
\begin{center}
\leavevmode
\epsfig{file=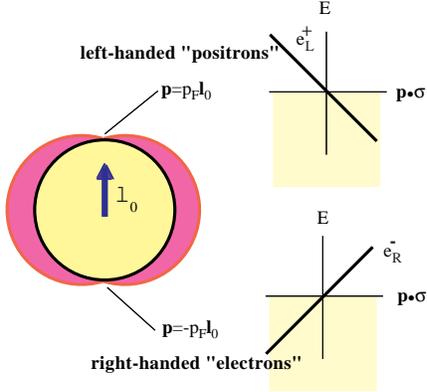,width=0.8\linewidth}
\caption[Analogy]
{a) Gap nodes and chiral fermions in $^3$He-A. The pink region illustrates how
the gap in the quasiparticle spectrum depends on position at the Fermi surface.
The gap has two nodes along the $\pm {\bf \hat l}$ directions which are shown
as the North and South poles of the Fermi sphere. The nodes occur at the North
pole only for those quasiparticles that have a negative   projection of the
Bogoliubov-Nambu spin along their momentum. Therefore they correspond to the
left-handed particles. Similarly, the node at the South pole applies only to
right-handed quasiparticles. In addition, the interactions of the
quasiparticles
near the gap nodes with the order  parameter is analogous to the interaction of
massless fermions with  background hyperelectromagnetic and gravitational
fields. The hypercharge of the right-handed quasiparticles is $-1$ and that of
left-handed quasiparticles is $+1$.
}
\label{Analogy}
\end{center}
\end{figure}

\noindent (i) The Cooper pairs in $^3$He-A have angular momentum
$\hbar$ and locally all pairs have their angular momentum aligned along a
direction ${\hat {\bf l}}$, which also indicates
(Fig.~\ref{Analogy}(a))
the direction to the gap nodes of Bogoliobov quasiparticles.
In particle physics language, the $^3$He-A vacuum is described by the uniform
expectation value of a unit vector field ${\hat{\bf l}}$ and the fermionic
energy levels have a zero mode for momenta in the direction of
${\hat{\bf l}}$. Close to the gap nodes these quasiparticles represent
chiral relativistic fermions: they are left-handed in the vicinity of the
north
pole
${\bf p} \approx + p_F{\hat{\bf l}}$,  where $p_F$ is the Fermi momentum,
and they are right-handed in the vicinity of the south pole where
${\bf p} \approx -p_F{\hat{\bf l}}$.

In general, the vector ${\hat{\bf l}}$ will not be absolutely uniform. For
example it may oscillate around some background distribution ${\bf {\hat
l}}_0$, which we further choose as uniform:
\begin{equation}
{\bf {\hat l}} = {\bf {\hat l}}_0 + \delta {\bf {\hat l}}({\bf r},t)\ .
\label{oscillating-l}
\end{equation}
For example, in the presense of counterflow, ${\bf w} ={\bf v}_n-{\bf
v}_s$, of the normal component of $^3$He-A liquid with respect to the
superfluid,  the constant part of the ${\bf {\hat l}}$-vector is oriented along
the flow: ${\bf {\hat l}}_0\parallel {\bf w}$.

The space and time dependence of $\delta {\hat{\bf l}}$
produces a force on   quasiparticles equivalent to the force of an
``hyperelectric field''
${\bf E}=k_F \partial_t \delta {\hat {\bf l}}$ (here $k_F= p_F/\hbar$) and a
``hypermagnetic field''
${\bf B}=k_F{\bf \nabla}\times \delta {\hat {\bf l}}$ acting on relativistic
massless fermions of unit charge $e=\pm 1$. Thus the analog of the  vector
potential ${\bf A}_H$ of the hyperelectromagnetic field is played by
$k_F \delta {\hat{\bf l}}$.

It is important for our consideration that the $^3$He-A liquid is anisotropic
in the same manner as a nematic liquid crystal. For the relativistic
fermions this means that their motion is determined by the  geometry
of an effective spacetime. In $^3$He-A this geometry is described by the
following meric tensor
\begin{equation}
g^{ik}= c_\perp^2 (\delta^{ik} - \hat l^i_0 \hat l^k_0) +
c_\parallel^2 \hat l_0^i
\hat l^k_0 ~~,~~ g^{00}=-1
\label{metric}
\end{equation}
The   quantities $c_\parallel= p_F/m$ and
$c_\perp= \Delta_0/p_F$ correspond to velocities of ``light'' propagating
along and transverse to  ${\hat{\bf l}}_0$, here $m$ is the mass of the $^3$He
atom and $\Delta_0$ is the amplitude of the gap.

In the presense of counterflow, ${\bf w} ={\bf v}_n-{\bf
v}_s$, of the normal component of $^3$He-A liquid with respect to the
superfluid,  the constant part of the ${\bf {\hat l}}$-vector is oriented along
the flow: ${\bf {\hat l}}_0\parallel {\bf w}$. In the vicinity of the
gap nodes, the energy of quasiparticles is Doppler shifted by the amount
${\bf p}\cdot{\bf w}\approx \pm p_F({\hat{\bf l}}_0\cdot{\bf w})$.
The counterflow therefore produces what would be an effective chemical
potential in particle physics. For right-handed particles, this is
$\mu_R= p_F({\hat{\bf l}}_0\cdot{\bf w})$ (Fig.~\ref{Analogy}(b)) and for
left-handed particles it is $\mu_L=-\mu_R$.

If the direction of the counterflow is chosen such that $\mu_R > 0$,
the quantities in $^3$He-A corresponding to the energy and number density
in eq. (\ref{ernr}), are the kinetic energy of the counterflow and the
${\bf{\hat l}}_0$-projection of its linear momentum,
${\bf P}=\hat\rho_n {\bf w}$
\begin{equation}
\epsilon_R\equiv {1\over 2}{\bf w} \hat\rho_n {\bf w}  ~~,~~ n_R\equiv{1\over
p_F}{\bf P}\cdot {\bf{\hat l}}_0 ~.
\label{he3ernr}
\end{equation}
Here $\hat\rho_n$ is the density of the normal
component, which in the anisotropic $^3$He-A is a tensor. The normal component
consists of the thermally activated quasiparticles and corresponds to the
system of chiral fermions.

To make sure that eq. (\ref{he3ernr}) is analogous to eq. (\ref{ernr}), let
us consider the
low-temperature limit $T\ll T_c$, where $T_c \sim \Delta_0$ is the superfluid
transition temperature. Then using eq. (3.94) of \cite{VollhardtWolfle}
for the density of the normal component of $^3$He-A and the $^3$He-A
equivalent of the chemical potential one obtains
\begin{equation}
\epsilon_R =
\approx {1\over 6} m k_F^3
{T^2\over \Delta_0^2}   ({\hat{\bf l}}_0\cdot {\bf w} ) ^2
\equiv {1\over 6} \sqrt{-g}T^2 \mu_R^2 ~~.
\label{lowternr}
\end{equation}
In the last equality an over-all constant appears to be the square
root of the determinant of the effective metric in $^3$He-A:
$\sqrt{-g}=1/c_\parallel c_\perp^2=mk_F/\Delta_0^2$. This makes
eq. (\ref{lowternr}) and  eq. (\ref{ernr})  absolutely equivalent.
The factor $\sqrt{-g}$ should also be
included in eq. (\ref{ernr})  as it is part of the  spatial volume element.
However, isotropy is assumed in the cosmological  scenario and so, in Cartesian
coordinates, the factor is equal to 1.

\noindent (ii)
The inhomogeneity which absorbs the fermionic charge, is represented by a
magnetic field configuration in real vacuum and by a $\delta {\hat{\bf
l}}$-texture  in $^3$He-A. However, eq. (\ref{anomaly}) applies in both cases,
if in $^3$He we use the identification ${\bf A} = k_F \delta {\bf {\hat l}}$.

Just as in the particle physics case, we now consider the instability
towards the production of texture due to the excess of chiral particles.
This instability can be seen by considering the energy of the inhomogeneous
texture on the background of the superflow. In the geometry of the superflow,
the gradient  contribution to the free energy of the
$\delta{\hat{\bf l}}$-texture is completely
equivalent  to the conventional energy of the hypermagnetic field
\begin{eqnarray}
F_{\rm magn}= {\rm ln}~ \left ( {\Delta_0^2\over T^2 }~ \right )
{{p_F^2v_F}\over {24\pi^2\hbar}}~
( \hat {\bf l}_0\times ({\bf\nabla} \times  \delta \hat {\bf l}))^2
\nonumber \\
~~\equiv
{{\sqrt{-g}}\over {4\pi}e_{eff}^2} g^{ij}g^{kl}F_{ik}F_{jl}~~.
\label{magenergy}
\end{eqnarray}
Here $F_{ik}= \nabla_i A_k -\nabla_k A_i $ and we have included the
effective anisotropic metric in eq. (\ref{metric}) appropriate for $^3$He-A.

It is interesting that the logarithmic factor in the gradient energy
plays the part of the running coupling
$e_{eff}^{-2} =(1/3\pi \hbar c)\ln (\Delta_0/T)$ in particle physics,
where $e_{eff}$ is the effective hyperelectric charge; while the
gap amplitude $\Delta_0$, plays the part of an ultraviolet cutoff
energy scale.
Now if one has the counterflow in $^3$He-A, or its equivalent --  an excess of
chiral charge produced by the chemical potential $\mu_R$ --  the anomaly
gives rise to an additional effective term in the magnetic energy,
corresponding to the interaction of the charge absorbed by the magnetic field
with the chemical potential. This effective energy term is:
\begin{eqnarray}
F_{CS}=(n_{R}-n_L)\mu_R &=&
{1\over 2\pi^2} \mu_R  {\bf A}\cdot ({\bf\nabla}\times {\bf A}) \nonumber \\
&=& {3\hbar\over 2m}\rho ({\hat{\bf l}}_0\cdot{\bf w})
(\delta{\hat{\bf l}}\cdot {\bf\nabla}\times \delta{\hat{\bf l}})~~,
\label{csenergy}
\end{eqnarray}
The right-hand side corresponds to the well known anomalous interaction
of the counterflow with the ${\hat{\bf l}}$-texture in $^3$He-A,
where $\rho$ is the mass density of $^3$He \cite{exotic} (the
additional factor of 3/2 enters due to nonlinear effects).

For us the most important property of this term is that it is linear in
the derivatives of
$\delta {\bf {\hat l}}$. Its sign thus can be negative, while its magnitude can
exceed the positive quadratic term in eq. (\ref{magenergy}). This leads  to the
helical instability towards formation of the inhomogeneous $\delta {\bf {\hat
l}}$-field. During this instability the kinetic energy of the
quasiparticles in the counterflow (analog of the energy sitting in the
fermionic
degrees of freedom) is converted into the energy of inhomogeneity
$ {\bf\nabla}\times
\delta{\hat{\bf l}}$, which is the analog of the magnetic energy of the
hypercharge field.

\begin{figure}[!!!t]
\begin{center}
\leavevmode
\epsfig{file=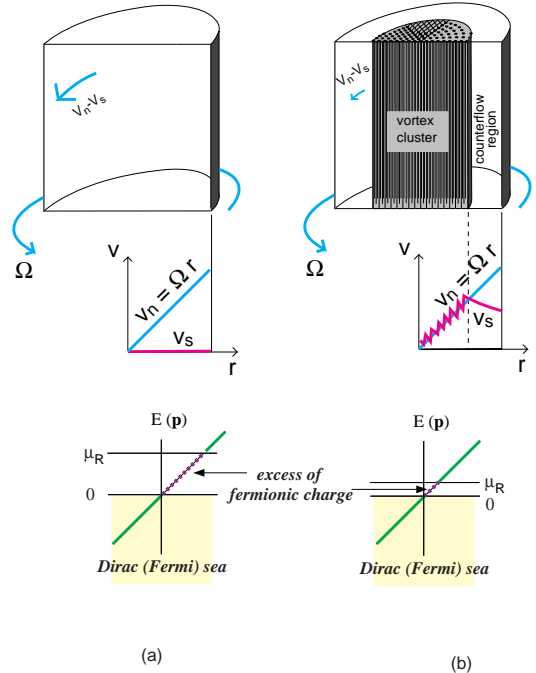,width=0.8\linewidth}
\caption[CounterflowAsCharge]
{a) ({\it top}) The vortex-free state in the vessel rotating with angular
velocity ${\bf \Omega}$ contains a counterflow,
${\bf w}= {\bf v}_n-{\bf v}_s \ne 0$,
since the average velocity of quasiparticles (the normal component)
is ${\bf v}_n={\bf \Omega}\times {\bf r}$ and is not equal to
the velocity of the superfluid vacuum, which is at rest,
${\bf v}_s=0$.  In the counterflow state, the quasiparticles have a
net momentum $\propto {\bf w}$.
({\it bottom})  In the presence of counterflow ${\bf w}$, the energy of
quasiparticles is Doppler shifted by the amount
${\bf p}\cdot{\bf w}\approx \pm p_F({\hat{\bf l}}_0\cdot{\bf w})$.
The counterflow therefore produces what would be an effective chemical
potential in particle physics. For right-handed particles, this is
$\mu_R= p_F({\hat{\bf l}}_0\cdot{\bf w})$ and for left-handed particles it
is $\mu_L=-\mu_R$. Thus the excess of quasiparticle momentum is
analogous to the excess of chiral right-handed electrons if their
chemical potential $\mu_R$ is nonzero.
(b) The excess of the fermionic
charge, corresponding to the excess of the quasiparticle momentum in the
rotating cryostat, is unstable towards the formation of a hypermagnetic
field. The latter corresponds to the ${\hat{\bf l}}$-texture. Further
development
of the instability gives rise to the periodic ${\hat{\bf l}}$-texture,  which
represents a periodic array of continuous vortices,
called the Chechetkin \cite{Chechetkin}, and, Anderson and Toulouse\cite{AT}
vortices (ATC vortices).  Formation of these
vortices decreases the counterflow (fermionic charge). Thus  a part of the
fermionic charge is transformed into hypermagnetic field. }
\label{CounterflowAsCharge}
\end{center}
\end{figure}

Until now we had a strong analogy between particle physics and
the $^3$He system. An important difference, however, arises from the
``mass of the hyperphoton'', which stabilizes the counterflow in $^3$He-A.
In the electroweak theory the hyperphoton is massless at high temperatures
while in $^3$He-A the defined ${\bf A}$ field has a mass for technical
reasons: to experimentally visualize the nucleation of the
``hypermagnetic field'' ${\bf B}$ we need to apply the NMR technique
\cite{Experiment}, which uses a real magnetic field, {\it i.e.} the
electromagnetic magnetic field ${\bf H}$.  In $^3$He-A such a field provides a
mass to the ``hypercharge gauge field'' ${\bf A}$, so that, in the presence
of ${\bf H}$, the instability occurs only above a critical value
$\mu_R^{cr}$ of the counterflow, which depends on the temperature $T$.

\begin{figure}[!!!t]
\begin{center}
\leavevmode
\epsfig{file=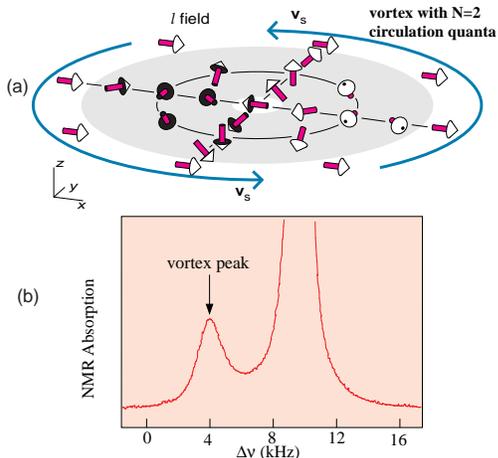,width=0.8\linewidth}
\caption[experiment]
{(a) The development of the counterflow
instability leads to the nucleation of the
ATC vortex. In the rotating vessel the
${\hat{\bf l}}$-vector forms a regular periodic structure. Each elementary
cell of the structure represents the ATC vortex with a quantum of
circulation of the superfluid velocity about the cell boundary $C$:
$\oint_C d{\bf r}\cdot {\bf v}_s=h/m$.
(b) The NMR signal from the  array of ATC vortices in the
container\cite{Parts}.
The position of the satellite peak indicates the type of the vortex, while the
intensity is proportional to the number of vortices in the cell.}
\label{experiment}
\end{center}
\end{figure}

\begin{figure}[!!!t]
\begin{center}
\leavevmode
\epsfig{file=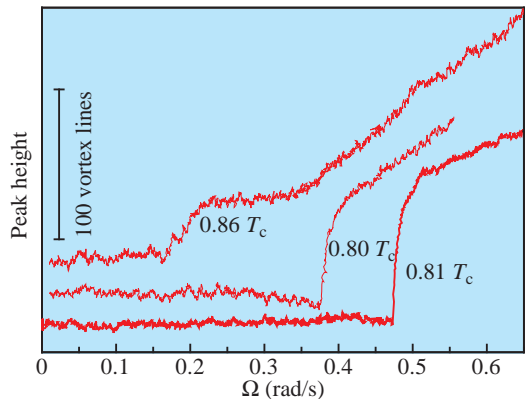,width=0.8\linewidth}
\caption[experiment2]
{Time dependence of the peak height of the continuous vortices. Initially
no vortices are present in the vessel. When the velocity of the counterflow
${\bf w}$ in the ${\hat{\bf l}}_0$ direction (corresponding to the chemical
potential $\mu_R$ of the chiral electrons) exceeds a critical value, an
instability takes place, and the container  becomes filled with the
${\hat{\bf l}}$-texture (analog of a hypermagnetic field) forming the vortex
array.}
\label{experiment2}
\end{center}
\end{figure}

When the helical instability develops in $^3$He-A, the final result is
the formation of a ${\hat{\bf l}}$-texture, which corresponds to the free
energy minimum in the rotating container.  This is a periodic ${\hat{\bf
l}}$-texture, where the elementary cell represents a so-called
Anderson-Toulouse-Chechetkin (ATC) continuous vortex  (see
Fig.~\ref{experiment}(a)). ATC vortices give rise to a characteristic
satellite peak in the $^3$He NMR absorption spectrum: their number is directly
proportional to the height of this peak (see Fig.~\ref{experiment}(b)).

By accelerating the container into rotation, starting  from the state with
vortex-free counterflow (ie. with fermionic charge, but no hypermagnetic
field),
we observe the helical instability as a sudden discontinuity, when the vortex
satellite is formed (see  Fig.~\ref{experiment2}). The peak height jumps from
zero to a magnitude which almost corresponds to the equilibrium vortex state,
which means that the counterflow is essentially reduced. Most of the
counterflow (fermionic charge) thus becomes converted into vortex texture
(magnetic field).

The treshold value $\Omega_c$ of the rotation velocity
$\Omega$, at which the helical instability occurs, determines the critical
value of the chemical potential $\mu_R^{cr}=\Omega_cRp_F$, where $R$ is the
radius of the vessel. The magnitude of $\mu_R^{cr}$, which we find from the
measurements \cite{Experiment},  is in good quantitative agreement with the
theoretical estimation of the mass of the ``hyperphoton'', determined by the
spin-orbit interaction in $^3$He-A. Thus we have modeled the formation of the
hypermagnetic field for different masses of the ``hyperphoton''. We have also
observed the flow instability in the limit when the ``hyperphoton'' has zero
mass: first the field ${\bf H}$ was reduced to zero, then the container was
accelerated to rotation at some velocity $\Omega$, and finally the rotating
state, which had been obtained in this way, was analyzed in the previous
manner in the NMR conditions. Since the critical velocity decreases with
decreasing magnetic field, the field depencence of the instability could be
worked out by this technique and the ``hypermagnetic field'' could be measured.
In this case $\mu_R^{cr}$ was essentially reduced.

Our result together with that in \cite{BevanNature} show that the chiral
anomaly is an important mechanism for the interaction of textures
(the analogue of the hypercharge magnetic fields and cosmic strings) with
fermionic excitations (analogue of quarks and leptons). We have thus
experimentally checked both processes which are induced by the anomaly: the
nucleation of fermionic charge from the vacuum in \cite{BevanNature} and the
inverse process of the nucleation of an effective magnetic field from the
fermion current.

{\bf Acknowledgements.}
This collaboration was supported by ESF. TV was supported
through a research grant from the Department of Energy, USA.

\pagebreak


\begin{thebibliography}{15}

\bibitem{exotic} G.E. Volovik,  Exotic properties of superfluid
$^3$He, World Scientific, Singapore-New Jersey-London-Hong Kong,
1992.

\bibitem{VolovikVachaspati}  G.E. Volovik and T. Vachaspati, Aspects of
$^3$He and the standard electroweak model, Int. J. Mod. Phys. {\bf B~10},
471--521 (1996).

\bibitem{BevanNature}  T.D.C. Bevan, A.J. Manninen, J.B. Cook, J.R. Hook,
H.E. Hall, T. Vachaspati and G.E. Volovik,
Momentogenesis by $^3$He vortices: an
experimental  analogue of primordial baryogenesis,  Nature,  {\bf 386},
689-692  (1997).

\bibitem{roberge} A. Roberge, ``Finite Density Effects in Gauge Theories'',
Ph. D. thesis, University of British Columbia (1989).

\bibitem{tvgf} T. Vachaspati and G. B. Field,
``Electroweak String Configurations with Baryon Number'',
{\it Phys. Rev. Lett.} {\bf 73}, 373 (1994); {\bf 74}, {\it Errata} (1995).

\bibitem{tvsintra} T. Vachaspati, ``Electroweak Strings, Sphalerons and
Magnetic Fields'', in the Proceedings of the
NATO Workshop on ``Electroweak Physics and the Early Universe'',
Sintra, Portugal (1994); Series B: Physics Vol. 338, Plenum Press,
New York (1994).

\bibitem{JoyceShaposhnikov}  M. Joyce, M. Shaposhnikov, Primordial
magnetic fields, right electrons, and the abelian anomaly, Phys.
Rev. Lett.,{\bf 79}, 1193-1196 (1997).

\bibitem{GiovanniniShaposhnikov} M. Giovannini and E.M.
Shaposhnikov,  Primordial
hyperagnetic fields and triangle anomaly, hep-ph/9710234.

\bibitem{Vilenkin} A. Vilenkin, ``Equilibrium Parity-Violating
Current in a Magnetic Field'', Phys. Rev. {\bf D22}, 3080 (1980).

\bibitem{Adler1969} S. Adler, Axial-vector vertex in spinor
electrodynamics, Phys. Rev. {\bf 177}, 2426--2438 (1969).

\bibitem{BellJackiw1969} J.S.Bell and R.Jackiw, A PCAC
puzzle:$\pi^0\rightarrow \gamma\gamma$ in the $\sigma$-model, Nuovo Cim.  {\bf
A60}, 47--61 (1969).

\bibitem{ThesisVollhardt} D. Vollhardt, ``Stability of supeflow and
related textural transformations in superfluid $^3$He'', PhD
Thesis, Hamburg, 1979.


\bibitem{VollhardtWolfle} D. Vollhardt, P. and P. W\"olfle,  The
superfluid phases of helium 3,  Taylor and Francis, London - New York -
Philadelphia, 1990.

\bibitem{Experiment}  V.M.H. Ruutu, J. Kopu, M. Krusius, U. Parts, B.
PlaŠais, E.V. Thuneberg, and W. Xu V.M.H. Ruutu, Critical Velocity of
Vortex Nucleation in Rotating Superfluid 3He-A,
Phys. Rev. Lett. {\bf 79}, 5058 (1997).


\bibitem{Chechetkin} V.R. Chechetkin, JETP,  {\bf 44}, 706 (1976).

\bibitem{AT} P.W. Anderson and G. Toulouse, Phys. Rev. Lett.,
{\bf 38}, 508 (1977).

\bibitem{Parts} \"U. Parts, J.M. Karim\"aki, J.H.
Koivuniemi,  M. Krusius, V.M.H. Ruutu, E.V. Thuneberg, and
G.E. Volovik, Phase diagram of vortices in superfluid $^3$He-A,
Phys. Rev. Lett. {\bf 75}, 3320--3323 (1995).


\end{thebibliography}
\end{document}